\documentclass[twocolumn,showpacs,preprintnumbers,amsmath,amssymb,prb]{revtex4}

 \def\XXint#1#2#3{{\setbox0=\hbox{$#1{#2#3}{\int}$}
     \vcenter{\hbox{$#2#3$}}\kern-.5\wd0}}

\def\fakebold#1{\relax\ifvmode\leavevmode\fi%
\ifmmode%
\setbox0=\hbox{$#1$}%
\else%
\setbox0=\hbox{#1}%
\fi%
\kern-.02em\copy0 \kern-\wd0%
\kern .04em\copy0 \kern-\wd0%
\kern-.0125em\raise.02em\box0%
}%

\usepackage{graphicx} \usepackage{dcolumn} \usepackage{bm}

\begin{document}

%\preprint{APS/123-QED}

%%%%%%%%%%%%%%%%%%%%%%%%%%%%%%%%%%%%%%%%%%%%%%%%%%%%%%%%%%%%%%%%%%%%%%%%%%%%%%%

\title{Exchange instabilities in electron systems: Bloch versus Stoner
Ferromagnetism}

\author{Ying Zhang} 
\author{S. Das Sarma} 
\affiliation{Condensed
  Matter Theory Center, Department of Physics, University of Maryland,
  College Park, MD 20742-4111}

\date{\today}

\begin{abstract}

We show that 2D and 3D electron systems with the long-range Coulomb
electron-electron interaction could develop ferromagnetic
instabilities due to strong exchange effects at low densities. The
critical densities in both 2D and 3D systems at which the magnetic
instability, which could either be of Stoner type (second-order) or of
Bloch type (first-order), are higher than the dispersion instability
critical density where effective mass at the Fermi surface diverges.
We discuss the theoretical as well as experimental implications of the
ferromagnetic instability at low electron densities, particularly in
low-disorder semiconductor-based two-dimensional systems.
\end{abstract}


\maketitle

%%%%%%%%%%%%%%%%%%%%%%%%%%%%%%%%%%%%%%%%%%%%%%%%%%%%%%%%%%%%%%%%%%%%%%%%%%%%%%%
\section{Introduction}
\label{sec:intro}

With the development of high mobility semiconductor based 2D electron
systems such as Si MOSFETs and GaAs HIGFETs, where extremely low
carrier densities and very high quality can be achieved, a number of
recent experiments has been carried out to measure fundamental
physical quantities such as magnetic susceptibility~\cite{vitkalov,
  kravchenko1, kravchenko2, kravchenko3, prus, shayegan1, shayegan2,
  shayegan3, shayegan4, zhu, pudalov, pudalov2, pudalov3} and
effective mass~\cite{tan} in low density 2D electron systems at low
temperatures ($\lesssim 100mK$). These experiments have triggered many
theoretical as well as experimental studies on 2D and 3D electron
systems. One important purpose of these studies is to obtain the
ground state phase diagram for such systems. It is well known that the
zero temperature 2D and 3D electron systems can be characterized by a
single dimensionless interaction parameter $r_s = E_{e-e} / E_K$
(where $E_{e-e}$ denotes the interaction energy and $E_K$ the kinetic
energy), which depends only on the density ($n$) of the system with
$r_s \propto n^{-1/2}~( n^{-1/3})$ in 2D (3D) systems.  For small
$r_s$, electron systems can be well described by Landau's Fermi liquid
theory. (Actually, there is an exception even to this as at
exceptionally low temperatures an interacting Fermi liquid undergoes a
Khon-Luttinger superconducting transition, which we ignore for our
purpose since it is of no physical relevance.) In the limit of large
$r_s$, the system tends to reduce its interaction energy $E_{e-e}$ at
the cost of higher kinetic energy by forming into an electron crystal
(the so called Wigner crystal~\cite{wigner}), which has been
established by Monte Carlo studies in both 2D~\cite{tanatar} (where
the crystallization transition happens at $r_s \sim 32 - 42$) and 3D
systems~\cite{ortiz} (where the transition happens at $r_s \sim 55 -
75$). How the system behaves in between the above mentioned two limits
of $r_s$ is not yet clear. It has been widely accepted, and also
suggested by Monte Carlo studies, that there may exist a ferromagnetic
phase (of either fully or partially polarized spins) in the
intermediate $r_s$ region. Many other theoretical studies for such
high $r_s$ value region typically start with a more or less arbitrary
assumption of a particular ground state symmetry of the system. Among
the various model of exotic interaction-driven electronic ground
states, charge or spin density wave states, various superconducting
states, glassy or clustered ground states have been much discussed in
the literature.

From a theoretical perspective, it is important to study the evolution
of the Fermi liquid state starting from the weakly interacting (small
$r_s$) regime. As $r_s$ increases, all the single particle properties
of the system are increasingly renormalized by interaction effects.
The question is whether the Coulomb interaction renormalization brings
about certain instabilities eventually at some large $r_s$, and hence
changes the ground state symmetry. The existence of degrees of freedom
related to spin and momentum makes it natural to consider the
possibility of a magnetic instability or an instability that is
related to the dispersion of a quasiparticle, which we call
``dispersion instability'' for short. In spite of a great deal of past
theoretical work investigating the magnetic instability~\cite{bishop,
  vyurkov, benenti, isihara1, isihara2, isihara3, shelykh1, shelykh2,
  overhauser, misawa, shastry, giuliani1, giuliani2, giuliani3, janak,
  doniach, coffey1, coffey2} and dispersion instability
possibilities~\cite{diverge, dispinst} in both 2D and 3D electron
systems using a large variety of different theoretical techniques, a
unified theoretical treatment of the ferromagnetic instability,
starting from the weak-coupling Fermi liquid ground state with
interaction effects introduced systematically, is still lacking. This
absence of a unified picture has left many unanswered important
questions. First, what is the order of any possible ferromagnetic
instability?  It could be a second-order transition (the so-called
Stoner instability) caused by the continuous divergence of the
susceptibility as the critical density or $r_s$-value is approached
from the weak-coupling side. Or it could be a first-order transition
(the so-called Bloch ferromagnetism) which could happen abruptly at a
specific $r_s$-value.  Both have been predicted and studied in the
literature, but their inter-relationship has not been clarified.
Second, what is the relationship between magnetic instability and
dispersion instability? Whether the possible divergence of the
interacting spin susceptibility $\chi^*$, which can be written as
$\chi^* = g^* m^*$, where $m^*(g^*)$ are the effective mass
($g$-factor) of the system, is caused by the divergence of $m^*$
(dispersion instability) or $g^*$, or both? In order to answer these
questions, we provide in this paper a comprehensive picture of the
Bloch ferromagnetic instability, Stoner ferromagnetic instability and
the dispersion instability of the 2D and 3D electron system using the
method of Random Phase Approximation (RPA)~\cite{book}.

The structure of the paper is the following: in Section \ref{sec:back}
we discuss briefly the background for Bloch and Stoner ferromagnetism;
in \ref{sec:theory} we describe our theoretical formalism; in
\ref{sec:results} we present and discuss our theoretical results for
ferromagnetic instability in interacting 2D and 3D quantum Coulomb
systems; we conclude in \ref{sec:con} with a discussion of
experimental implications and related open questions.

%%%%%%%%%%%%%%%%%%%%%%%%%%%%%%%%%%%%%%%%%%%%%%%%%%%%%%%%%%%%%%%%%%%%%%%%%%%%%%%
\section{Background}
\label{sec:back}

The possibility of a density driven ferromagnetic transition in an
interacting electron system was first suggested by Bloch~\cite{bloch}
more than 75 years ago. Bloch's basic idea, essentially a Hartree-Fock
mean field theory, remains fundamentally valid even today. The idea is
that at high density the electron system would be paramagnetic in
order to optimize the kinetic energy cost (which is high in a
high-density quantum fermionic system) whereas at low density the
system should spontaneously spin-polarize itself into a ferromagnetic
ground state in order to optimize the exchange energy arising from the
Pauli principle and Coulomb interaction. For an electron gas (in a
positive jellium background) it is a straightforward exercise to write
down the total Hartree-Fock energy per-particle as a sum of the
non-interacting kinetic energy and the (Fock) exchange energy due to
unscreened Coulomb interaction at $T=0$ as
\begin{eqnarray}
\label{eq:fock3D}
{E \over n} = {E_{KE} \over n} + {E_{ex} \over n}
&=& {0.55 \over r_s^2} \left[ (1 + \zeta)^{5/3} 
+ (1 - \zeta)^{5/3} \right] \nonumber \\
&-& {0.23 \over r_s} \left[(1 + \zeta)^{4/3} + (1 - \zeta)^{4/3} \right]
\end{eqnarray}
for 3D system, and
\begin{eqnarray}
\label{eq:fock2D}
{E \over n} = {0.50 \over r_s^2} (1 + \zeta)^2 
- {0.30 \over r_s} \left[ (1 + \zeta)^{3/2} + (1 - \zeta)^{3/2} \right]
\end{eqnarray}
for 2D systems, where $n = n_{\uparrow} + n_{\downarrow}$ is the total
number density of electrons; $\zeta = (n_{\uparrow} -
n_{\downarrow})/n$ is their spin-polarization (or magnetism) density;
$a_B^3 n = ({4 \over 3} \pi r_s^3)^{-1}$ and $a_B^2 n = (\pi
r_s^2)^{-1}$ define the dimensionless interaction parameters $r_s$ in
2D and 3D respectively (with $a_B = \hbar^2/(m e^2)$, the Bohr
radius); and the energy is measured in Rydberg units (i.e. $e^2/a_B$).
It is easy to see that the above Hartree-Fock energy expressions lead
to a first-order ferromagnetic transition (the ``Block
ferromagnetism'') at $r_s = r_B$ where $r_B \simeq 5.45$ (3D) and $r_B
\simeq 2$ (2D), i.e. $E(\zeta = 1)$ ferromagnetic state is lower
(higher) in energy than $E(\zeta = 0)$ paramagnetic state for $r_s >
(<) r_B$. We refer to such energy-difference-driven abrupt (first
order) transition as Bloch ferromagnetism in the rest of this paper.

The Stoner ferromagnetic instability~\cite{stoner} refers to the
divergence of the spin susceptibility and hence a second-order
continuous magnetic phase transition from a paramagnetic ($r_s <
r_{St}$) weak-coupling side to a ferromagnetic side ($r_s > r_{St}$).
The simplest model to consider is, following Stoner's original work, a
zero-range delta-function like interaction of strength `$I$' (a
constant in momentum space) between the electrons, leading to an
interacting static long-wavelength spin susceptibility (in the
dynamical Hartree-Fock approximation) given by $\chi^*/\chi = (1-D_0
I)^{-1}$, where $D_0 \equiv D(E_F)$ is the electronic density of
states at the Fermi energy. This immediately leads to the Stoner
criterion for ferromagnetic instability defined by a divergent
$\chi^*/\chi$ when $1 - D_0 I = 0$. Since the 2D density of states is
a density-independent constant, this instability criterion does not
lead to a meaningful condition in 2D unless we arbitrarily define $I$
to be the Coulomb interaction strength at Fermi wavevector, whence the
Stoner instability criterion leads to unphysically low $r_s$ values
for the ferromagnetic instability given by $r_{St} \simeq 0.7$ (2D)
and $r_{St} \simeq 1.5$ (3D), which are absurdly small $r_s$-values
and are unrealistic. Of course, in real electron systems the
electron-electron interaction is the long-range Coulomb interaction,
and therefore the simple Stoner instability criterion, defined by $D_0
I = 1$ where $I$ is an effective short-range interaction strength, is
inapplicable. But the basic idea of the ferromagnetic Stoner
instability, defined by a continuous divergence of $\chi^*(r_s)$ as
$r_s$ is increased, still applied. We refer to the ferromagnetic
transition defined or characterized by a divergence of the interacting
susceptibility as the Stoner instability.

In real electron liquids, the exchange-only Hartree-Fock approximation
considered above for the Bloch instability is inadequate because
correlation effects (i.e.  energy contributions beyond Hartree-Fock)
are known to be extremely important, and must be included in the
energetic considerations.  Similarly, the interacting susceptibility
must be calculated for the real Coulomb interaction in the system,
{\em not} for a hypothetical zero-range interaction, in order to
obtain a better estimate of the Stoner instability criterion. In the
rest of this paper we consider both Bloch and Stoner ferromagnetic
instabilities using better many-body approximations for Coulomb
electron liquids, namely RPA.

%%%%%%%%%%%%%%%%%%%%%%%%%%%%%%%%%%%%%%%%%%%%%%%%%%%%%%%%%%%%%%%%%%%%%%%%%%%%%%%
\section{Theory}
\label{sec:theory}

\begin{figure} [htbp]
\centering \includegraphics[width=3in]{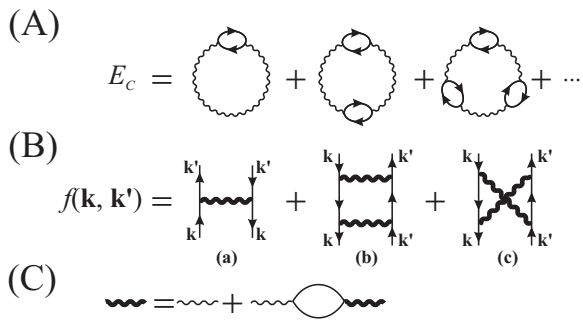}
\caption{The RPA Feynman diagram for: (A) the Coulomb interaction contribution
  to the ground state energy; (B) Landau's interaction function; (C)
  Dynamically screened interaction. The circles are polarization
  bubbles, the thin wiggly lines are the bare Coulomb interaction, and
  the solid lines the noninteracting electron Green's function.}
  \label{fig:feynman1}
\end{figure}

In this work we follow the notation of Refs.~\onlinecite{diverge,
  dispinst, mass}.  Within RPA~\cite{diverge, dispinst, book, mass,
  rice, hedin, hubbard}, the Coulomb contribution to the ground state
energy of a jellium electron system with long-range Coulomb
interaction can be denoted by the Feynman diagrams shown in
Fig.~\ref{fig:feynman1}A. The quasiparticle energy is then obtained by
$E_{\bf k} = {\delta E_G / \delta n_{\bf k}}$, where $n_{\bf k}$ is
the distribution function at momentum ${\bf k}$.  The second order
derivative of the total ground state energy is referred to as Landau's
interaction function: $f({\bf k}, {\bf k'}) = {\delta^2 E_G / \delta
  n_{\bf k} \delta n_{\bf k'}}$, represented by the Feynman diagram
shown in Fig.~\ref{fig:feynman1}B. Graphically, taking the $n_{\bf k}$
variational derivative of a quantity simply means that one cuts one
solid line of the Feynman diagram and takes the external momentum and
frequency to be on-shell (i.e.  $\omega = {\bf k}^2/2m - E_F$ with $m$
the band electron mass and $E_F$ the Fermi energy).  We emphasize that
the RPA as shown in Fig.~\ref{fig:feynman1} necessarily implies that
this on-shell self-energy approximation is used for calculating the
quasiparticle energy dispersion $E_{\bf k}$ and Landau's interaction
function $f({\bf k}, {\bf k'})$ since all energy and momenta in
Fig.~\ref{fig:feynman1} correspond to the noninteracting system. Thus
RPA self-energy approximation necessarily implies an on-shell
approximation (Fig.~\ref{fig:feynman1}A) as emphasized by
Rice~\cite{rice} a long time ago. Solving the full Dyson equation
using the off-shell renormalized energy, as is sometimes done in the
literature, would be completely inconsistent within RPA as it would
incorrectly mix various perturbative orders.

Following Hubbard's notation~\cite{hubbard}, the ground state energy
per particle $E_G /N$ with $N$ the particle number can be written as
${ E_G / N} = {E_K / N} - v(0) /2 + {E_C / N}$, where $E_K$ is the
kinetic energy, $v(0) = \int v_q {d^d q / (2 \pi)^d}$ is the
interaction energy at zero separation with the bare Coulomb
interaction $v_q = 2 \pi e^2 / q$ for 2D and $v_q = 4 \pi e^2 /q^2$
for 3D, and $E_C$ is the Coulomb contribution to the ground state
energy (both exchange and correlation) which can be denoted as in
Fig.~\ref{fig:feynman1}A. Note that the singularities in $v(0)/2$ and
$E_C / N$ cancel out with each other.  In the 2D system, it is easy to
show~\cite{hubbard} that RPA leads to
\begin{eqnarray}
\label{eq:EG2}
{E_G \over N} &=& {E_F \over 2} + {E_{ex} \over N}
\nonumber \\
&& \!\!\!\!\!\!\!\!
+ {16 E_F \over g_s \pi} \int_0^\infty x d x \int d u
[ \ln \epsilon(x, u) - \epsilon(x, u) ],
\end{eqnarray}
where $\epsilon(q, \omega)$ is the dynamical dielectric function,
$g_s$ is the spin degeneracy ($g_s=2$ for paramagnetic states and $g_s
= 1$ for fully spin-polarized state), $E_{ex}$ is the exchange part of
the ground state energy. Note that in Eq.~(\ref{eq:EG2}) we subtracted
a term $\epsilon(q, \omega)$ from $\ln[\epsilon(q,\omega)]$ in order
to handle the ultraviolet divergence in the integration.  Similarly
for 3D we have
\begin{eqnarray}
\label{eq:EG3}
{E_G \over N} &=& {3 \over 5} E_F + {E_{ex} \over N}
\nonumber \\
&& \!\!\!\!\!\!\!\!
+ {48 E_F \over g_s \pi} \int_0^\infty x^2 d x \int d u
[ \ln \epsilon(x, u) - \epsilon(x, u) ].
\end{eqnarray}

It is convenient to convert all the expressions in terms of the
dimensionless units $r_s$. The definition of $r_s$ is as $ r_s =
1/(\alpha k_F a_B)$ where $k_F$ is the Fermi momentum and $a_B = 1 /
(m e^2)$ is the Bohr radius. In 2D $\alpha = {\sqrt{g_s} / 2}$ and in
3D $\alpha = ({2 g_s / 9 \pi})^{1/3}$.  Also we choose $\hbar = 1$
throughout, which makes wavevector and momentum (as well as energy and
frequency) equivalent.  In these notations, it is easy to show that
$E_{ex} / N = - 8 \alpha r_s / (3 \pi) E_F$ for 2D, and $E_{ex} / N =
- 3 \alpha r_s / (2 \pi) E_F$ for 3D. In the actual calculation, the
integration in Eqs.~(\ref{eq:EG2}) and (\ref{eq:EG3}) can be performed
on either real or imaginary axis. By examining the $r_s$ and $g_s$
dependence of the ground state energy, we study the Bloch magnetic
instability of the electron system. The integrals in
Eqs.~(\ref{eq:EG2}) and (\ref{eq:EG3}) are the correlation
contributions not included in our Hartree-Fock considerations of Bloch
ferromagnetic in Sec.~\ref{sec:back}. Note that $g_s=1(2)$ corresponds
to $\zeta = 1(0)$ in Sec.~\ref{sec:back}.

%*****************************************************************************

\begin{figure}[htbp]
  \centering
  \includegraphics[width=3in]{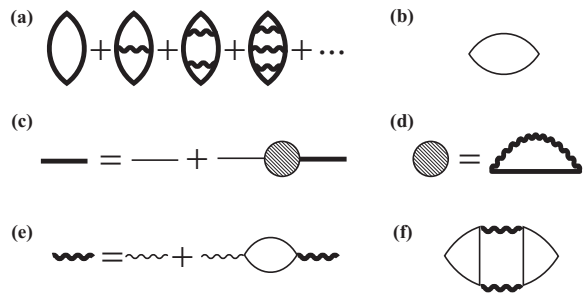}
  \caption{(a) The ladder-bubble series for the interacting
    susceptibility with the bold straight line the interacting Green's
    function and the bold wavy line the dynamically screened
    interaction; (b) the noninteracting susceptibility; (c) the
    Dyson's equation for the interacting Green's function in terms of
    the noninteracting Green's function and the self-energy; (d) the
    self-energy in the leading-order expansion in the dynamical
    screening; (e) the Dyson's equation for the dynamically screened
    interaction in terms of the bare Coulomb interaction (thin wavy
    lines) and the polarization bubble; (f) a charge fluctuation
    diagram which does not contribute to spin susceptibility.}
  \label{fig:chifeynman}
\end{figure}

We investigate the Stoner instability by calculating the magnetic
susceptibility $\chi^*$ within RPA which is represented by the Feynman
diagram showed in Fig.~\ref{fig:chifeynman}. Direct calculation of
these diagrams turns out to be difficult for the long ranged Coulomb
interaction. However, at $T=0$, Landau showed that $\chi^*$ can be
equivalently expressed as the following equation~\cite{rice}:
\begin{equation}
\label{eq:chi}
{\chi \over \chi^*}={m \over m^*} + \varpi \int f_e (\theta) do,
\end{equation}
where $\chi$ is the Pauli spin susceptibility, $f_e(\theta)$ =
$f_e({\bf k}, {\bf k'})$ with ${\bf k}$ and ${\bf k'}$ on-shell; ${\bf
  k}^2/2m = {\bf k'}^2/2m (= E_F)$ is the exchange Landau's
interaction function, $\theta$ is the angle between ${\bf k}$ and
${\bf k'}$, $do$ the element of solid angle along ${\bf k'}$ times
$\cos\theta$ in 3D and $d\theta$ in 2D, and $\varpi=1/(2\pi)^2$ in 2D
and $\varpi=k_F/(2\pi)^3$ in 3D.  Similarly, the Landau theory
expression for the effective mass $m^*$ is~\cite{rice}
\begin{equation}
\label{eq:m}
{m \over m^*}=1 - \varpi \int f(\theta) do.
\end{equation}
In Eq.~(\ref{eq:chi}), $f_e(\theta)$ is the exchange part of the
Landau's interaction function, which is represented by
Fig.~\ref{fig:feynman1}B(a). This spin independent term is responsible
for the difference between the ratio $\chi/\chi^*$ and
$m/m^*$~\cite{rice}. An equivalent, and easier way to derive the
effective mass is through calculating quasiparticle self-energy and
obtaining its momentum derivative. The self-energy within RPA can be
written as~\cite{diverge}
\begin{equation}
\label{eq:E0}
\Sigma({\bf k}, \omega) = - \int {d^d q \over (2 \pi)^d} 
\int {d \nu \over 2 \pi i} {v_q \over \epsilon({\bf q}, \nu)} 
G_0 ({\bf q} + {\bf k}, \nu + \omega), 
\end{equation}
where $d = 2$ or $3$ is the dimension of the system, and
\begin{equation}
\label{eq:G0}
G_0({\bf k}, \omega) =
{1 - n_F(\xi_{\bf k}) \over \omega - \xi_{\bf k} + i 0^+ } +
{n_F(\xi_{\bf k}) \over \omega - \xi_{\bf k} - i 0^+ } 
\end{equation}
is the bare Green's function, where $n_F$ is the Fermi distribution
function, $\xi_{\bf k} = k^2 /(2m) - E_F$. It is shown~\cite{dispinst}
that the integration along real axis in the expression of self-energy
(Eq.~(\ref{eq:E0})) can be deformed onto imaginary axis, which avoids
the singularities along the real axis and makes the integration
easier.  The contour deformation also breaks the expression of
self-energy into separate terms that correspond respectively to
contributions from the spin-dependent and spin-independent part of the
Landau's interaction function shown in Fig.~\ref{fig:feynman1}, and is
very useful for us to derive the expression for susceptibility as
shown below. The expression of the real part of the self-energy can
then be written as
\begin{eqnarray}
\label{eq:ReE}
&& \!\!\!\!\!\!\!\!
\mbox{Re}~\Sigma({\bf k}, \omega) =- \int {d^d q \over (2 \pi)^d} v_q 
\Theta(2 m \omega + k_F^2 - |{\bf q - k}|^2) \nonumber \\
&& + \int {d^d q \over (2 \pi)^d} 
v_q \mbox{Re}~{1 \over \epsilon({\bf q}, \xi_{\bf q - k} - \omega)} 
\nonumber \\
&&~~~\cdot \Big[\Theta(2 m \omega + k_F^2 - |{\bf q - k}|^2) 
- \Theta(k_F^2 - |{\bf q - k}|^2) \Big]\nonumber \\
&& - \int {d^d q \over (2 \pi)^d} \int {d \nu \over 2 \pi}
v_q \left[ {1 \over \epsilon({\bf q}, i \nu)} -1 \right]
{1 \over i \nu + \omega - \xi_{{\bf q} + {\bf k}}}.
\end{eqnarray}
The effective mass is derived from the expression of the real part of
the quasiparticle self-energy by $m/m^* = 1 + (m/k_F) {d \over d k}
\mbox{Re}~\Sigma({\bf k}, \xi_{\bf k}) |_{k = k_F}$~\cite{rice}.
Combining this with Eq.~(\ref{eq:chi}), we have
\begin{equation}
\label{eq:chi2}
{\chi \over \chi^*}=1 + \varpi \int f_e (\theta) do
+ {m \over k_F} {d \over d k} 
\mbox{Re}~\Sigma({\bf k}, \xi_{\bf k}) |_{k =
  k_F}.
\end{equation}
It is not difficult to show that the second term of Eq.~(\ref{eq:ReE})
accounts for the contribution from the spin-independent exchange
Landau's interaction function $f_e({\bf k}, {\bf k'})$
(Fig.~\ref{fig:feynman1}B(a)), and therefore the term $\varpi \int f_e
(\theta) do$ in Eq.~(\ref{eq:chi2}) exactly cancels the momentum
derivative of the second term in the self-energy Eq.~(\ref{eq:ReE}).
Hence the expression of $\chi / \chi^*$ only contains contributions
from the $k$ derivatives of the first and third term in
Eq.~(\ref{eq:ReE}). After converting all the expressions in terms of
the dimensionless parameter $r_s$, and using $2k_F$, $4 E_F$, $2m$ as
the momentum, energy, and mass units, we obtain the expression for the
2D magnetic susceptibility as
\begin{eqnarray}
\label{eq:sus2D}
&& \!\!\!\!\!\!\!\!\!\!
{\chi \over \chi^*} = - {2 \alpha r_s \over \pi} 
 + { \sqrt{2} \alpha r_s \over \pi} 
\int_0^\infty x^2 d x \int_0^\infty d u 
\left[{1 \over \epsilon(x,  iu)} -1 \right] 
\nonumber \\ 
&&~~~\times \left[A \sqrt{1 + A/R} - B \sqrt{1-A/R}\right] R^{-5/2},
\end{eqnarray}
where $A = x^4 - x^2 - u, B = 2 x u, R = \sqrt{A^2 + B^2}$.
Similarly for 3D we have
\begin{eqnarray}
\label{eq:sus3D}
{\chi \over \chi^*} &=& - {\alpha r_s \over \pi} 
+ { \alpha r_s \over 2 \pi^2} \int_0^\infty d x \int_0^\infty d u 
\left[ {1 \over \epsilon(x,  iu)} -1 \right]
\nonumber \\ 
&& ~~~~~~ \times
\left[ \ln ( F/G ) -  2C / F + 2D / G \right],
\end{eqnarray}
where $C = 1 - q, D = 1 + q, F = C^2+u^2, G=D^2+u^2$. Note that in the
expressions of the dielectric function $\epsilon(x, u)$, $x = q / (2
k_F)$ and $u = \omega / (4 E_F)$.

%%%%%%%%%%%%%%%%%%%%%%%%%%%%%%%%%%%%%%%%%%%%%%%%%%%%%%%%%%%%%%%%%%%%%%%%%%%%%%%
\section{Results}
\label{sec:results}

\begin{figure}[htbp]
\centering \includegraphics[width=3.5in]{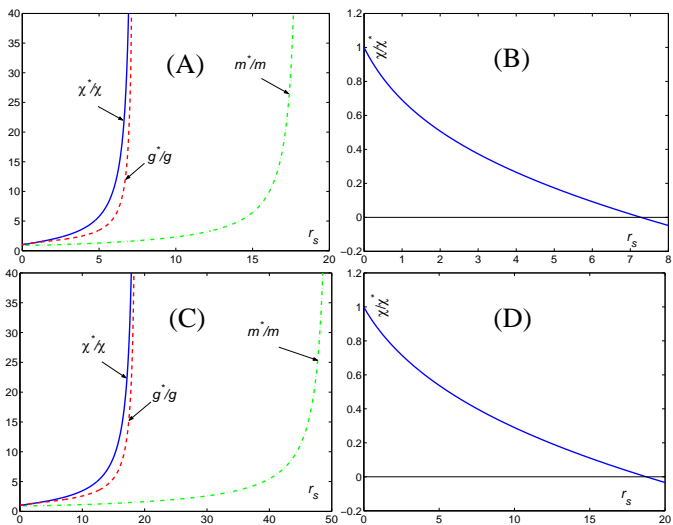}
  \caption{(A) and (C): Calculated renormalized spin susceptibility 
    $\chi^*/\chi$, effective mass $m^*/m$ and $g$-factor $g^*/g$. For
    2D system, $\chi^*$ and $g^*$ diverge at $r_s \sim 7.3$ while
    $m^*$ diverges at $r_s \sim 18.1$. For 3D system, $\chi^*$ and
    $g^*$ diverges at $r_s \sim 18.7$ while $m^*$ diverges at $r_s
    \sim 49.9$. (B) and (D): inverse susceptibility shows $\chi^*$
    diverges at $r_s \sim 7.3$ for 2D and $18.7$ for 3D systems. Note
    that the $\chi^*$ and $g^*$ are calculated for paramagnetic
    systems ($g_s = 2$) while $m^*$ are for ferromagnetic systems
    ($g_s = 1$).}
\label{fig:sus} 
\end{figure}

In Fig.~\ref{fig:sus} we present the calculated magnetic
susceptibility as a function of $r_s$ for both 2D and 3D systems,
together with the calculated $g$-factor $g^* = \chi^*/m^*$ and the
effective mass $m^*$. It is clear from Fig.~\ref{fig:sus} that both 2D
and 3D systems experience Stoner ferromagnetic instabilities,
characterized by the divergence of magnetic susceptibility as the
density decreases.  It is important to note that this Stoner
instability (i.e. divergence of $\chi^*$) does not arise from an
effective mass divergence since the $m^*$ divergence happens at much
lower densities. In other words, $g^*$ and $m^*$ both diverge, with
the divergence of $g^*$ occurring at a lower $r_s$ value. For the 2D
system, $\chi^*$ and $g^*$ diverges at $r_s \sim 7.3$ while $m^*$
diverges at $r_s \sim 18.1$.  For the 3D system, $\chi^*$ and $g^*$
diverge at $r_s \sim 18.7$ while $m^*$ diverges at $r_s \sim 49.9$.

We emphasize that both $m^*$ and $g^*$ divergences actually happen
independently and are completely unrelated phenomena. On the other
hand, it is also worth mentioning that the $g^*$ divergence does have
some quantitative effect on the $m^*$ divergence. After $g^*$
diverges, the system becomes a ferromagnetic liquid, and the momentum
distribution is different and the Fermi energy increases. This change
results in a small increase in the critical $r_s$ value where $m^*$
diverges. In fact for $g_s = 2$ paramagnetic systems, $m^*$ diverges
at $16.1$ for 2D and $47.8$ for 3D~\cite{diverge}, in contrast to $g_s
= 1$ case where $m^*$ diverges at $18.1$ for 2D and $49.9$ for 3D. But
this effect is rather small and is of no particular significance.

\begin{figure}[htbp]
\centering \includegraphics[width=3.5in]{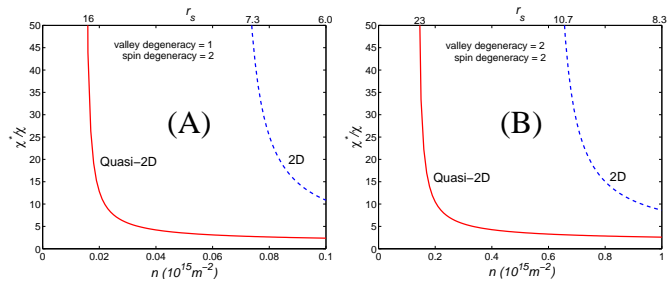}
  \caption{Quasi-2D effects on magnetic susceptibility divergence. 
    (A): GaAs quantum well system. (B): Si-inversion layer. }
\label{fig:finiteW} 
\end{figure}

In real experimental systems, the value of $r_s$ at which $\chi^*$
diverges should be influenced by many factors (even within our RPA
many-body approximation scheme). Here we consider the valley
degeneracy $g_v$ and the finite width effect on the magnetic
susceptibility divergence in semiconductor-based realistic 2D electron
systems. The effect of $g_v$ is exactly the same as the effect of
$g_s$ on the system, and therefore can be easily incorporated. For the
finite width effect, we introduce a form factor to the Coulomb
interaction, following the standard procedure described in detail in
Refs.~\onlinecite{sus, ando}. Using appropriate semiconductor
parameters, we obtain the susceptibility in GaAs quantum wells and
Si-inversion layers, plotted in Fig.~\ref{fig:finiteW}. It is clear
from Fig.~\ref{fig:finiteW} that multi-valley degeneracy and finite
width both suppress the divergence of the susceptibility
renormalization, and make the critical $r_s$ value for $\chi^*$
divergence considerably larger than the strict 2D results.

\begin{figure}[htbp]
\centering \includegraphics[width=3.5in]{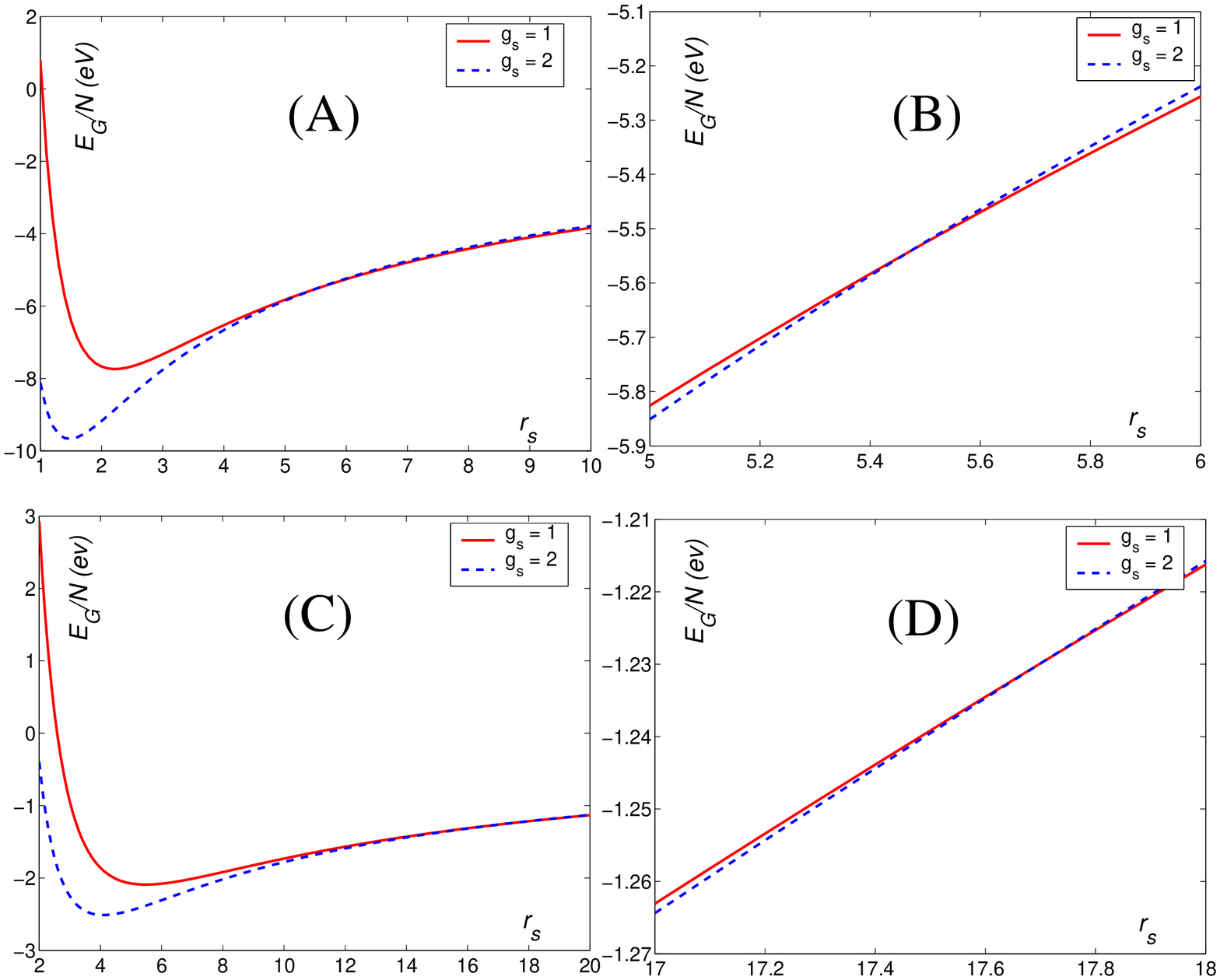}
  \caption{(A) and (C): the RPA ground state energy per particle 
    for polarized ($g_s = 1$) unpolarized ($g_s = 2$) cases as a
    function of $r_s$ for 2D and 3D systems, calculated using
    Eq.~(\ref{eq:EG2}) and Eq.~(\ref{eq:EG3}).  (B) and (D): details
    around the magnetic instability at $r_s \sim 5.5$ for 2D and
    $17.8$ for 3D systems.}
\label{fig:EG} 
\end{figure}

The magnetic susceptibility is a thermodynamic Fermi-surface property.
As mentioned before, another way of studying the magnetic instability
is to compare the ground state energy of the system for polarized and
unpolarized states at different electron densities. This is the Bloch
ferromagnetism discussed in Sec.~\ref{sec:back}. Our results
(Fig.~\ref{fig:EG}) of RPA ground state energy for fully polarized and
non-polarized electron states (using Eqs.~(\ref{eq:EG2}) and
(\ref{eq:EG3})) in both 2D and 3D electron systems show similar
characteristic. When $r_s$ is very small (or electron densities high),
both systems prefer non-polarized paramagnetic states.  As $r_s$
increases to a certain critical value ($r_s \sim 5.5$ for 2D and $r_s
\sim 17.8$ for 3D) the ground state energy for the fully polarized
electron state actually becomes lower than the non-polarized states.
This clearly indicates that the system undergoes a Bloch type
ferromagnetic instability due to the Coulomb electron-electron
interaction in a low density 2D electron systems. Note that the
critical $r_s$ for the Bloch instability is substantially higher in
the RPA theory (increasing from $2$ to $5.5$ in 2D and $5.45$ to
$17.8$ in 3D) than in the Hartree Fock theory due to the inclusion of
correlation energy.

Comparing Figs.~\ref{fig:sus} and \ref{fig:EG} we conclude that, at
least within our well-defined RPA ring-diagram many-body approximation
scheme, the sequence of instabilities (as density decreases) the
theory predicts for both 2D and 3D electron liquids is the following:
Bloch ferromagnetism (i.e. an abrupt first order magnetic transition)
at $r_s = 5.5$ (2D); $17.8$ (3D); Stoner ferromagnetism characterized
by a continuous divergence (i.e. a second order magnetic transition)
of the interacting $g$-factor and of the susceptibility at $r_s = 7.3$
(2D), $18.7$ (3D); the dispersion instability associated with the
continuous divergence of the quasiparticle effective mass at $r_s =
18.1$ (2D), $49.9$ (3D). Of course, comparing first and second order
transitions is not particularly meaningful since their origins are
fundamentally different, and a first order transition may always
preempt a second order transition as seems to happen in Coulomb
electron liquids with Bloch ferromagnetism always happening (both in
2D and 3D) at a slightly lower $r_s$ values ($5.5$ versus $7.3$ in 2D,
and $17.8$ versus $18.7$ in 3D) although the difference in the
critical $r_s$ values for the two transitions (less than $10\%$ in 3D
and about $25\%$ in 2D) is sufficiently small so that both Bloch and
Stoner ferromagnetism remain of experimental interest.

We also note that, by definition, $\chi^* \equiv g^* m^*$ (and $\chi =
g m$), and therefore the divergence of the interacting susceptibility
could be caused either by a divergence $g$-factor or a diverging
effective mass. This issue has been much discussed and
debated~\cite{vitkalov, kravchenko1, kravchenko2, kravchenko3,
  pudalov, pudalov2, pudalov3} in the recent experimental literature
on 2D semiconductor-based electron systems, where low-density
divergence of both 2D susceptibility and effective mass has been
reported. All we can say is that our theoretical results are only
consistent with the susceptibility divergence as arising from the
divergence of the interacting $g$-factor, not the effective mass,
since the $g$-factor divergence occurs at much lower $r_s$-values,
$r_s \sim 7.3$ ($18.7$) in 2D (3D) for $g^*/g$ divergence versus $r_s
\sim 18.1$ ($49.9$) in 2D (3D) for $m^*/m$ divergence. We add that in
realistic quasi-2D semiconductor system (our Fig.~\ref{fig:finiteW})
the susceptibility (as well as effective mass) divergence occurs at
substantially higher $r_s$ values due to the considerable softening of
the Coulomb interaction from its strict 2D form due to the
finite-width effect.

%%%%%%%%%%%%%%%%%%%%%%%%%%%%%%%%%%%%%%%%%%%%%%%%%%%%%%%%%%%%%%%%%%%%%%%%%%%%%%%
\section{Discussion}
\label{sec:con}

Both Bloch and Stoner instabilities imply that 2D and 3D electron
systems interacting via the long range Coulomb interaction undergo a
$T=0$ ferromagnetic quantum phase transition from a high-density
paramagnetic state to a low-density ferromagnetic state either through
a first order (Bloch) transition or a second-order (Stoner) transition
(with a continuous divergent susceptibility) at a critical
$r_s$-value.  Given that the critical $r_s$ value(s) for the
transition(s) we obtain within our RPA many-body theory is rather
large (i.e. $r_s \gg 1$), we do not expect our predicted $r_s$
parameter for ferromagnetic transitions in 2D and 3D electron systems
to be reliable. But the basic trends, such as the sequence of
instabilities (i.e. Bloch followed by Stoner followed by the
dispersion instability with diverging mass as the density is lowered)
or the suppression of the transition to much lower densities in {\em
  quasi}-2D systems, should be valid, in general. Indeed quantum phase
transitions predicted by RPA (or for that matter even by the simpler
Hartree-Fock approximation) are always found to occur in the numerical
quantum Monte Carlo (QMC) simulation albeit at higher $r_s$ values.
This is certainly true for electron liquid ferromagnetic
instabilities.  For 3D system, QMC simulation predicts that the system
undergoes a possibly second-order phase transition and electron spins
become partially polarized at $r_s \sim 60$ by Ref.~\onlinecite{zong}
or $r_s \sim 15 - 25$ by Ref.~\onlinecite{ortiz}, and as $r_s$
increase to $r_s \sim 100$ by Ref.~\onlinecite{zong} or $r_s \sim
35-45$ by Ref.~\onlinecite{ortiz}, the system becomes a fully
polarized ferromagnetic state.  For 2D case, QMC simulation
predicts~\cite{attaccalite} that as density approaches $r_s \sim 25$,
the system undergoes a first-order transition into a fully
spin-polarized ferromagnetic state.  Theoretically it is possible to
obtain the spin susceptibility through calculating the second
derivative of the ground state energy with respect to $\zeta$ at
$\zeta = 0$. However, in reality too much error is introduced when
obtaining the susceptibility this way in QMC simulations, and
therefore it is difficult to ascertain the order of the ferromagnetic
transition in QMC numerical simulations -- our RPA theory predicts the
first-order Bloch instability to occur first as the density is being
lowered. Another thing to be noticed about QMC simulation results is
that the ferromagnetic transition $r_s$ value predicted by different
groups differs by a large factor from each other (see, for example,
Ref.~\onlinecite{ortiz} and Ref.~\onlinecite{zong}), which shows the
large amount of error introduced by such simulations due to the small
energy difference between spin polarized states and spin unpolarized
states (the two density dependent energy curves are almost parallel
when they cross each other) and different choices of trial wave
functions.

Much has been written about the validity of the RPA many-body
approximation at low carrier densities ($r_s > 1$). We have little to
add to this issue beyond the detailed discussion we already provided
in our recent publications~\cite{mass, diverge}. We want to emphasize
that, although RPA is exact in the $r_s \to 0$ limit, it is by no
means a theory based on an $r_s$ expansion -- it is a self-consistent
field theory based on an expansion in the dynamically screened
interaction which should be qualitatively valid for {\em all} $r_s$
below the Wigner crystallization of the electron liquid. In fact, RPA
is found to be quantitatively valid in 3D systems~\cite{rice} at
metallic densities ($r_s \sim 3 - 6$) and in 2D systems for $r_s$ up
to $10-15$ where comparison with experiment has been carried
out~\cite{hwang}. Often the error in the calculation arising from
other effects (e.g.  finite temperature~\cite{finiteT}, finite
quasi-2D width, band structure, etc.) turn out to be larger than that
included in the RPA approximation, and therefore improvement beyond
RPA (short of a full-fledged QMC calculation) becomes meaningless.
One can try to ``improve'' upon RPA by including local field
corrections to the dynamical electron polarizability (i.e. bare bubble
of RPA) which, in some crude manner, simulates the incorporation of
higher-order vertex corrections in the theory. But such local field
corrections are uncontrolled, and probably inconsistent, since many
diagrams in the same order are typically left out. We are therefore
unconvinced that the inclusion of local field corrections in the
theory is necessarily an improvement on RPA. The great conceptual
advantage of RPA is that it is a well-defined approximation that is
both highly physically motivated (i.e. dynamical screening) and
theoretically exact in the high-density ($r_s \to 0$) limit. Attempted
improvement upon RPA through the arbitrary inclusion of local field
correction may neither be theoretically justifiable nor more reliable.
Keeping these caveats in mind we have carried out our exchange
instability calculations using the Hubbard local field
corrections~\cite{hubbard}, and we find no qualitative changes from
the RPA results presented in this paper. The critical $r_s$ values for
the occurrence of the ferromagnetic instability change somewhat in the
presence of the local field correction, but this is a result without
any significance since the precise values of critical $r_s$ are
expected to be not particularly accurate in any of these theories. The
fact that the basic qualitative conclusions about the various
instabilities do not change in the presence of local field corrections
demonstrates the qualitative robustness of our RPA-based results.
Another point to note is that the fact that RPA predicts the existence
of divergence of certain physical quantities at certain critical
densities does not necessarily imply that RPA or Fermi liquid theory
fails at that density. For example, RPA predicts the divergence and
then negative values for the compressibility~\cite{book} at densities
higher than ferromagnetic transition densities, and negative
compressibility has indeed been observed in
experiments~\cite{eisenstein} on 2D electron systems as predicted by
RPA calculations. Of course the {\em total} compressibility of a
system cannot be negative, but just the electronic part of the
compressibility can be negative as predicted by RPA and as is
routinely observed in 2D electron systems for $r_s > 3$. There can be
no doubt that if 3D electron systems with large enough $r_s$ values
($r_s > 6$ according to RPA~\cite{book}) are found they would
routinely have negative electronic compressibility as well! It is
certainly true that RPA becomes a progressively poorer approximation
as density decreases and perhaps detailed QMC calculations should be
carried out to test the validity of our RPA-based predictions
presented in this paper.

In discussing possible experimental implications of our results, we
note the great recent experimental interest in the literature on the
possibility of a density-driven ferromagnetic transition in
semiconductor-based 2D carrier systems where very low carrier
densities ($r_s \lesssim 10$) can be achieved in rather high-quality
samples~\cite{vitkalov, kravchenko1, kravchenko2, kravchenko3, prus,
  shayegan1, shayegan2, shayegan3, shayegan4, zhu, pudalov, pudalov2,
  pudalov3, tan}. There are recent experimental claims~\cite{vitkalov,
  kravchenko1, kravchenko2, kravchenko3} of the observation of a low
density susceptibility divergence in Si MOSFETs for $r_s \sim 7 - 10$.
Although the experimental susceptibility behavior as a function of
density (or, $r_s$) looks similar to our theoretical results in
Figs.~\ref{fig:sus} and \ref{fig:finiteW}, we are skeptical about the
significance of this agreement. There are several reasons for our
skepticism. First, the experimental claimed susceptibility divergence
occurs at far too high a density ($r_s \sim 7$) compared with the
theory where we find the RPA susceptibility divergence in realistic Si
MOSFETs (Fig.~\ref{fig:finiteW}) to be occurring at $r_s \approx 23$.
This RPA predictions for critical $r_s (\sim 23)$ is most likely the
lower bound -- any real susceptibility divergence is expected to occur
at higher $r_s$ values ($r_s > 23$). Second, the experimental
divergence of $\chi^* / \chi \equiv g^* m^* / g m$ has been
claimed~\cite{kravchenko1, kravchenko2, kravchenko3} to be arising
from an effective mass divergence, {\em not} the $g$-factor divergence
as we find in our theory. Our theoretical effective mass divergence,
in fact, occurs at a critical $r_s$ more than twice as large as the
corresponding $\chi^*$ critical $r_s$. In fact, our {\em quasi}-2D
effective mass divergence~\cite{diverge, dispinst} occurs for $r_s >
40$! Third, there has been no experimental evidence for the existence
of a low-density ferromagnetic phase such as hysteresis, remanence,
etc. If there is indeed a ferromagnetic transition, one should be able
to observe ferromagnetic behavior at densities below the ferromagnetic
critical density (i.e. for $r_s$ values larger than the point of
$\chi^*$ divergence). No such direct ferromagnetic behavior has ever
been observed in a low-density 2D system casting serious doubts on the
claims of the observation of a 2D ferromagnetic transition. A very
recent extremely careful and detailed measurement of 2D susceptibility
in a high-mobility n-GaAs system~\cite{zhu} finds no divergence in
$\chi^*/\chi$ up to $r_s \approx 12$, calling into question the
earlier claims of susceptibility divergence in Si MOSFETs at lower
$r_s$ values. In addition, a direct thermodynamic
measurement~\cite{prus} of 2D susceptibility in Si MOSFETs also does
not find a ferromagnetic instability. What is clear is the observed
strong enhancement of $\chi^*/\chi$ (and $m^*/m$) as a function of
increasing $r_s$ which is consistent with our theoretical findings.
But the actual existence of a low-density electron liquid
ferromagnetic transition has not be established experimentally in our
opinion.

Finally we discuss some of the earlier literature that is of relevance
to our work. Our calculation of the ground state energy for polarized
and unpolarized states partially confirms the numerical results of
Rajagopal {\it et.  al.}~\cite{rajagopal}, who also considered the
possibility of partially polarized states, and found that for certain
range of electron densities, the system prefers a partially polarized
ferromagnetic state in the 3D system. Similar results were also
derived using the Quantum Monte Carlo method~\cite{ortiz, zong}. In
this paper we did not present our calculation results (which confirms
the results of Ref.~\cite{rajagopal}) for the Bloch instability
associated with the partial spin-polarization because this will not
help our understandings of the relation between Stoner, Bloch, and
dispersion instabilities, which is the main purpose of this paper. In
3D, partially spin polarized states is preferred energetically for the
density region of $r_s$ between $\sim 14$ and $\sim 18$, which is
right before the fully polarized ferromagnetic region as the density
is decreased~\cite{rajagopal}. This suggests that our understanding of
the relation between the three kinds of instabilities will not be
affected by the consideration of partially spin-polarized states.  In
2D systems, partial spin-polarization does not occur~\cite{rajagopal,
  zong, ortiz}, and thus this issue does not arise at all for our 2D
calculations, which is the main focus of our work. For the magnetic
susceptibility, there have been earlier RPA calculations in
2D~\cite{giuliani1, giuliani2, giuliani3, giuliani4, giuliani5} and
3D~\cite{shastry}, and QMC calculations in 2D~\cite{attaccalite}. Only
Shastry~\cite{shastry} predicted a susceptibility divergence in 3D
systems, and our results confirm his conclusion. No previous work
considered the inter relations, among Bloch instability, Stoner
instability and dispersion instability. Our work is the only work in
the literature connecting all these density-driven electron liquid
instabilities within one coherent theoretical framework.

This work is supported by NSF, DARPA, ONR, LPS, and ARO.

%%%%%%%%%%%%%%%%%%%%%%%%%%%%%%%%%%%%%%%%%%%%%%%%%%%%%%%%%%%%%%%%%%%%%%%%%%%%%%%

\bibliography{maginst}

\end{document}